\documentclass[10pt,conference]{IEEEtran}
\usepackage{cite}
\usepackage{amsmath,amssymb,amsfonts}
\usepackage{algorithmic}
\usepackage{graphicx}
\usepackage{textcomp}
\usepackage{xcolor}
\usepackage{listings}

\begin{document}

\title{Let's Ask Students About Their Programs, Automatically}

\author{\IEEEauthorblockN{Teemu Lehtinen}
\IEEEauthorblockA{
\textit{Aalto University}\\
Espoo, Finland \\
teemu.t.lehtinen@aalto.fi}
\and
\IEEEauthorblockN{Andr\'e L. Santos}
\IEEEauthorblockA{
\textit{Instituto Universit\'ario de Lisboa (ISCTE-IUL)}\\
Lisboa, Portugal \\
andre.santos@iscte-iul.pt}
\and
\IEEEauthorblockN{Juha Sorva}
\IEEEauthorblockA{
\textit{Aalto University}\\
Espoo, Finland \\
juha.sorva@aalto.fi}
}

\maketitle

\begin{abstract}
Students sometimes produce code that works but that its author does not comprehend. For example, a student may apply a poorly-understood code template, stumble upon a working solution through trial and error, or plagiarize. Similarly, passing an automated functional assessment does not guarantee that the student understands their code. One way to tackle these issues is to probe students’ comprehension by asking them questions about their own programs. We propose an approach to automatically generate questions about student-written program code. We moreover propose a use case for such questions in the context of automatic assessment systems: after a student’s program passes unit tests, the system poses questions to the student about the code. We suggest that these questions can enhance assessment systems, deepen student learning by acting as self-explanation prompts, and provide a window into students’ program comprehension. This discussion paper sets an agenda for future technical development and empirical research on the topic.
\end{abstract}


\section{Introduction}

Students’ solutions to program-writing assignments are typically checked for correctness and, perhaps, other qualities such as style, efficiency, or test coverage. The results of this manual or automatic assessment are often accepted as a partial assurance that the learner has reached some intended learning outcomes. Most automatic assessment systems \cite{Keuning2018AExercises} attempt to provide rapid, constructive feedback on errors, so that students can fix their programs and learn from their mistakes.

Instructors would like students to design programs and write code that the students themselves understand. Reality bites, however. We, like many others, have seen learners search for and copy poorly understood code from textbooks and internet forums. Sometimes students fiddle with example code through trial and error until it happens to work, or are guided by teaching assistants or automatic assessment to the correct solution without reaching understanding. (Automatic assessment may exacerbate bad study habits such as trial-and-error and starting late.~\cite{Auvinen2015HarmfulAssessment})

Learners’ programs may then feature code that they do not properly comprehend: they cannot trace their program’s execution and may have a shaky grasp of the principles behind the code. Plagiarism is another reason for submitting code without understanding. Similar problems may also stem from legitimate collaboration with student peers. 

Whatever the underlying reasons, successful task performance is not the same as successful learning, and a student with a program that works is not the same as a student who understands why and how the program works. Program comprehension must be acknowledged as an educational objective and that objective should be addressed in pedagogy.

Our work addresses a dual issue: 1)~for deeper learning, students should reflect on the code that they write and ensure they understand it; 2)~the quality of assessment is limited by a sole focus on the resultant program at the expense of understanding. 

The second issue is problematic not only for teachers who wish to ensure that students do not get excessive credit for code they do not understand, but also for the learners themselves. For example, automatic assessment that focuses on the end product alone may lull learners into a false sense of understanding and lead them to miss out on important learning outcomes---a sort of \emph{unproductive success}~\cite{Kapur2016ExaminingLearning}.

In order to explore students' understanding and stimulate reflection, teachers can engage students in dialogue about their code and ask questions about it (see, e.g,~\cite{Brennan:NewCTFrameworks, Garcia:Questions}). However, when there are many students and even more programs, it is unfeasible to maintain such one-on-one human dialogue. This is the case especially in large introductory courses, such as those at Aalto University, where the rapid provision of feedback at scale is only possible because of automated assessment.

\section{Article Goals and Structure}

We propose that useful questions about student-written programs can be created automatically. These \emph{questions about learners' code} (QLCs) can then be posed to the same student who wrote the code. The driving question that we investigate is this:

\vspace{2mm}
\noindent \emph{What opportunities and challenges are there in automatically generating questions (QLCs) for students to answer about their own programs?}
\vspace{2mm}

\noindent We explore this question by connecting it to the research literature and through critical argumentation. This is a ``discussion paper.’’ That is, our present purpose is to put forward the idea of QLCs, to discuss its foundations in computing education research and the learning sciences, and to identify opportunities and challenges in implementing and adopting automatically generated QLCs. 

As we explore these questions, we also sketch out some use cases for QLCs. As an example use case, after a learner submits a program for automatic assessment and it passes all or most unit tests (or comparable checks of correctness), the system may ask one or more questions about the program.

The work presented here is theoretical, and does not yet feature a concrete software system to support QLCs or an empirical evaluation thereof. Instead, we outline future research on tool development and empirical evaluations of QLCs.

The rest of this article is structured as follows. Section~III introduces the concept of QLCs. Section~IV reviews the related work that forms the theoretical backdrop for QLCs, motivates this research, and informs QLC design. Section~V discusses, in broad terms, the technical problem of generating QLCs automatically and outlines a plausible process for approaching this problem. Section~VI identifies use cases for QLCs and considers some different forms that QLCs may take. Section~VII notes several limitations of the QLC idea. Section~VIII sets down an agenda for constructive and empirical research on QLCs, and Section~IX briefly concludes the article.

\section{Questions about Learners' Code}

\subsection{Definition}

\noindent We define QLCs as follows.

\begin{enumerate}
\item They are questions about program code\\that a student has written;
\item they refer to concrete constructs or patterns\\in the student's program; and
\item they are posed to the student themselves\\by a computer. 
\end{enumerate}

\noindent \emph{Automatic QLCs} have an additional characteristic:

\begin{enumerate}
\item [4)] They are automatically generated from\\an analysis of the student's code.
\end{enumerate}

\noindent An automatic QLC is not a canned, teacher-created question that is specific to a single programming assignment that students work on. Instead, we envision that teachers can affect which types of QLCs are generated for their students and, if they wish, configure question generation differently for different assignments. In this article, we focus on automatic QLCs. Hereafter, when we discuss QLCs, we mean automatic QLCs unless otherwise specified.

Many kinds of QLCs are possible. As an example, Figure~\ref{fig:program} presents a solution to a programming assignment; let us imagine this solution is from a learner. The goal of the assignment is to provide practice on program writing in general and recursion in particular. Figure~\ref{fig:questions} lists a few plausible QLCs for this code.

Even when QLCs are automatically \emph{generated}, their \emph{assessment} may or may not be automatic. Some QLCs have answers that are relatively simple to assess automatically, such as multiple-choice or single-value questions. On the other hand, QLCs may also have open-ended answers that are self-, teacher-, or peer-assessed. Abstract examples of both kinds of questions appear in Table~\ref{tab:absquestions}. 

\lstset{
  basicstyle=\footnotesize\ttfamily,        
  breakatwhitespace=false,         
  breaklines=true,                 
  captionpos=b,                    
  extendedchars=true,              
  frame=single,                    
  language=Java,                 
  keywordstyle=\bf,
  showspaces=false,                
  showstringspaces=false,          
  showtabs=false,                  
  tabsize=1,                       
  xleftmargin=.5em
}

\begin{figure}[t]
    \textbf{Task: Write a recursive function to find the smallest character in a String.}
\begin{lstlisting}
static char smallest(String word) {
  return smallestFrom(word, 0);
}

static char smallestFrom(String word, int index) {
  if(index == word.length() - 1) {
    return word.charAt(index);
  }
  else {
    char current = word.charAt(index);
    char rest = smallestFrom(word, index + 1);
    return current < rest ? current : rest;
  }
}
\end{lstlisting}
    \caption{An example task and a solution in Java.}
    \label{fig:program}
\end{figure}

\begin{figure}[t]
    \begin{enumerate}
        \item You wrote two functions. Which of those are recursive?
        \item What are the parameter names of your function {\tt smallestFrom}?
        \item How deep does the call stack grow when executing \texttt{smallest("ABBA")}?
        \item When executing \texttt{smallestFrom("ACDC", 0)}, which character is assigned to \texttt{rest} during the \emph{second} invocation of \texttt{smallestFrom}?
        \item Which of the following best describes the role of your variable \texttt{rest}?\footnotemark[1]
    \end{enumerate}
    \caption{Example QLCs for the code in Figure~\ref{fig:program}.}
    \label{fig:questions}
\end{figure}

\subsection{Motivation}

QLCs may engage learners to reflect on their code and their programming knowledge, thus enhancing learning. They may prompt additional practice on program comprehension, set in the (presumably) familiar context of the learner's own program. 

QLCs extend the assessment of student-created programs beyond functionality and style to program comprehension. By looking into the relationship between the learner and their program, QLCs complement other forms of assessment that focus on the program alone.

If collected, learners' answers to QLCs could be valuable to teachers and researchers who wish to explore students' programming knowledge. QLCs might also have uses in discouraging or detecting plagiarism. In the sections that follow, we  elaborate on these potential uses and benefits of QLCs.

\footnotetext[1]{\emph{Roles of variables} are common patterns of variable use~\cite{Sajaniemi2005AnProgramming} that can be taught to novices and identified automatically through static analysis~\cite{Santos2018EnhancingAnalysis}. For example, a \emph{gatherer} variable gradually accumulates a result by combining multiple inputs, and a \emph{most-wanted holder} tracks the highest, smallest, or otherwise most appropriate value among various candidates.}

\section{Related Work}

The subsections below relate our work on QLCs to four extant threads of research: student learning of code reading, knowledge elaboration through self-explanation, automatic generation of questions for computing education, and automated tutoring systems, respectively.

\subsection{Learning to Read Programs}

\subsubsection{Student Difficulties}

Even after taking a programming course, learners frequently have fragile program-reading skills, and many struggle to trace or explain code~\cite{Lister2004AProgrammers, Simon11}. Moreover, learners are sometimes averse to tracing their code even when it might help~\cite{Cunningham:RationalesForTracingAvoidance, Cunningham:IAmNotAComputer}.

Fuller et al.~\cite{Fuller2007DevelopingTaxonomy} discuss the learning trajectories of novice programmers in terms of an adaptation of Bloom's taxonomy. They cite evidence for some students getting stuck at an \emph{Apply/Remember} stage, where the student relies on trial and error to imitate the use of a construct without understanding. 

Salac and Franklin~\cite{Salac:IfTheyBuildIt} showed that, in a primary-school context, students' use of constructs is not a reliable indicator of understanding those constructs. Similarly, in a higher-education context, Kennedy and Kraemer~\cite{Kennedy:StudentReasoning} found that there were students who could produce a program that works but were unsure about their own code and did not grasp the underlying concepts.

QLCs target these difficulties by prompting students to reason about and explain the code they have produced. We propose that QLCs could prompt students’ progress from surface-level imitation to deeper understanding, and partially assess that progress.

\subsubsection{Types of Programming Knowledge}

Schulte's \emph{Block Model} of program comprehension~\cite{Schulte:BlockModel, Izu2019FosteringTrajectories} captures how understanding a program requires different kinds of knowledge---knowledge about the program text, about its dynamic behavior, and about the program's purpose; moreover, knowledge is required at different scales ranging from individual ``atoms'' to combinations of constructs to the full program. Table \ref{tab:blockmodel} summarises the Block Model. 

\begin{table*}[th]
  \caption{The Block Model (paraphrased from~\cite{Schulte:BlockModel})}
  \label{tab:blockmodel}
  \centering
  \begin{tabular}{|p{0.07\textwidth}|p{0.31\textwidth}|p{0.28\textwidth}|p{0.24\textwidth}|}
    \hline
    \textit{~} & \textbf{Text} & \textbf{Execution} & \textbf{Function} \\
    \hline
    \textbf{Atom} & language elements & elements' behavior & elements' purpose \\
    \hline
    \textbf{Block} & syntactically or semantically related elements & a ``block's'' behavior & a ``block's'' purpose; program subgoal \\
    \hline
    \textbf{Relational} & connections between ``blocks''; e.g., method calls & flow between ``blocks''; e.g., call sequences & integration of subgoals \\
    \hline
    \textbf{Macro} & entire program & the program's behavior & the program's purpose \\
    \hline
  \end{tabular}
\end{table*}

\noindent We suggest that useful QLCs can be generated at various levels of the Block Model, as illustrated by the examples in Table~\ref{tab:absquestions}. 

Understanding how code works is not valuable for its own sake only; the skill is required when writing, debugging, and extending programs~\cite{Izu2019FosteringTrajectories}. Research suggests that ability to trace code in detail and explain what given code accomplishes tends to precede ability to write comparable code oneself~\cite{Lopez:ReadingTracingWriting}. In any case, writing code that you can read is, by definition, predicated on code-reading ability. We hypothesize that QLCs can support the growth of program-comprehension skills and strengthen the relationship between them and program-writing skills.

\subsubsection{Pedagogies}

Recognizing the importance of reading code, scholars have explored pedagogies where learners practice reading given code or are explicitly taught strategies for tracing~\cite{LuxtonReilly:CS1Review}. For example, researchers have examined comprehension-before-writing pedagogies~\cite{Xie:TheoryOfInstruction}, invited learners to predict, investigate, and eventually modify the behavior of code~\cite{Sentance:PRIMM}, and asked students to visually simulate given programs~\cite{Sorva:DocThesis}.

The present work is loosely related, given that our goal is also to promote students’ understanding of existing code. However, our work differs materially in that QLCs do not involve teacher-provided code and are an extension of code-writing practice rather than a wholly distinct learning activity. QLCs are meant to deepen and assess learners' engagement with the code that they authored, and to do so automatically. We see QLCs as complementary---rather than alternative---to code-reading practice on given code.

Some tracing-based pedagogies emphasize the need for students to trace their own programs. For example, Hertz and Jump~\cite{Hertz:TraceBasedTeaching}, who taught students to use a diagrammatic notation for tracing code, noted that their code-writing problems always require students to trace the code they wrote. QLCs share the goal of promoting students’ understanding of their own code, but instead of graphical tracing, we propose to ask students to answer automatically generated questions of various kinds.

\subsection{Self-Explanation}

To \emph{self-explain}~\cite{Bisra:SelfExplReview} is to generate explanations for yourself---rather than a teacher or a peer---as you engage in a learning activity such as problem solving, reading some text, or studying an example. It is often but not necessarily internal, i.e., not expressed aloud or in writing. Self-explanations may explain conceptual content, justify the selection of solution steps, or otherwise elaborate on the learning activity. (Self-explanation is thus not a separate learning activity but a cognitive process that accompanies an activity; it may occur before, during, and/or after the activity.) 

Successful learners tend to generate better, more principled self-explanations, but many learners fail to do so~\cite{Atkinson:LearningFromExamples, Kwon:SelfExplProgramming}. Fortunately, teachers and environments can encourage self-explanation, and there is strong evidence (with caveats) for improved learning from \emph{self-explanation prompts} (SEPs)~\cite{RittleJohnson:ConstraintsOnSelfExpl, Bisra:SelfExplReview}; moreover, learners can be taught to self-explain better~\cite{Atkinson:LearningFromExamples, Bielaczyc:TrainingSelfExplanation}. SEPs are typically designed manually, but work on automatic prompt generation is underway~\cite{Bisra:SelfExplReview}.

The correlation between successful self-explanation and task performance has been replicated in programming contexts, as has the impact of self-explanation training~\cite{Kwon:SelfExplProgramming, Bielaczyc:TrainingSelfExplanation, Margulieux:TeachingSelfExpl}.
Moreover, research suggests that explicit subgoal-labeling promotes better self-explanations of example programs and thereby helps learners see structural similarities between examples and improves transfer~\cite{Margulieux:TeachingSelfExpl, Morrison:CuriousLoops}. Asking learners to annotate given code with comments~\cite{Vieira:SelfExplComments} or elaborate on their attempts to solve code-construction puzzles~\cite{Fabic:MobilePythonTutorActivities} are promising programming-specific SEPs.

Our work on QLCs overlaps that on SEPs; by promoting reflection, QLCs might have some of the same benefits. Depending on how QLCs are phrased, contextualized, and assessed, they may be considered as self-explanation prompts, assessment items, or both.

\begin{table*}[t]
  \caption{Example Questions from QLC Templates. (Simplified for brevity.)}
  \vspace{-2mm}
    \centering
  \label{tab:absquestions}
  \begin{tabular}{|p{0.66\textwidth}|p{0.11\textwidth}|p{0.15\textwidth}|}
    \hline
    \textbf{Template Question} & \textbf{Answer~Type} & \textbf{Block~Model~level~\cite{Izu2019FosteringTrajectories}} \\
    \hline
    Which of the following are variable names in your function?
    & multiple choice & atom--text \\
    \hline
    A loop starts on line [N]. Enter the number of the last line inside this loop.
    & single value & block--text \\
    \hline
    Line [N] uses a variable. Enter the line number where that variable is declared.
    & single value & relational--text \\
    \hline
    What is assigned to variable [V] on line [N] when executing function [F] on expression [E]?
    & single value & atom--execution \\
    \hline
    During the program execution, how many iterations are performed by the loop starting in line [N]?
    & single value & block--execution \\
    \hline
    Which of the following best describes the role of your variable [V]?
    & multiple choice & relational--execution \\
    \hline
    How deep does the call stack grow when executing function [F]?
    & single value & relational--execution \\
    \hline
    Describe the purpose of the condition on line [N].
    & open-ended & atom--function \\
    \hline
    Justify your choice of name [V] for the variable declared on line [N]---do you have a better suggestion?
    & open-ended & block--function \\
    \hline
    Select the part of your program that is responsible for [X].
    \emph{(The question could be generated from a subgoal annotated onto a model solution.)}
    & select in code & block--function \\
    \hline
    Explain, in your own words, the purpose of the loop that begins on line [N], and
    how that loop helps method [M] accomplish its task.
    & open-ended & relational--function \\
    \hline
    Here is a little example program that has some similarities with yours. Select the part of your program that serves a similar purpose as the highlighted code in the example. 
    & select in code & relational--function (across programs)\\
    \hline
  \end{tabular}
  \vspace{3mm}
\end{table*}

\subsection{Automatic Question Generation}

There is earlier work on automatically generating questions for programming education. For example, Zavala and Mendoza~\cite{Zavala2018OnExercises} designed exercise templates, each of which could be automatically instantiated into hundreds of specific variants of the same exercise, which vary in specific values and variable names. Thomas et al.~\cite{Thomas2019StochasticQuestions} automatically generated small programs accompanied by multiple-choice questions that prompted students to trace the generated code. This line of work is primarily motivated by the need to provide many fresh questions for students to practice on, and/or as a precaution for plagiarism. Our approach shares the general notion of automatically generated questions but differs fundamentally in that we do not aim to generate variants of the same exercise but to ask questions about existing code produced by the learner.

Outside of computing education, researchers have applied natural-language processing to text documents with the aim of automatically generating questions for learners~\cite{Lindberg2013, Labutov2015}. These questions have the potential to work as SEPs during independent study of the texts. The questions are open-ended with no well-defined correct answer; such questions are difficult to assess automatically or at scale, however. Our work has a related goal in automatically generating questions that may prompt self-explanation; we explore this idea in a programming context, generate the questions from learner-produced content, and look into automatic assessability.

\subsection{Automated Tutors for Programming}

\subsubsection{Intelligent Tutoring Systems}

Intelligent tutoring systems are designed to provide similar automated support as one-to-one human tutoring might. Over the years, many intelligent tutoring systems have been built for programming education. Some of these systems employ both automatically assessed programming exercises and questions to guide students towards set learning goals.~\cite{Crow2018}

As an example, the ITEM/IP tutoring system tests students' knowledge by asking questions that require the students to trace given example programs until they arrive to an answer~\cite{Brusilovsky1992}. QLCs expand on that idea by using the code that the student wrote as an input to generating comparable as well as other types of questions about the properties of the program.

\subsubsection{Automatic Assessment Systems}

Automated systems for assessing student programs are popular not only to grade student work at scale but also to generate detailed, rapid feedback that aims to improve that work and thereby to tutor the student. Feedback generation requires detailed static and dynamic analysis as well as program transformations~\cite{Keuning2018AExercises}. We propose to apply the previously researched program-analysis techniques in QLCs generation.

By providing feedback on programming exercises, automatic assessment systems generally aim to guide students while they are working towards fulfilling the functional requirements of a programming assignment. QLCs, on the other hand, prompt student to explain properties of a program artefact that may already have fulfilled some functional criteria---they are potentially relevant also to students who have submitted a functionally correct solution on the first attempt. We see QLCs as a promising extension to automated assessment systems, as discussed in Section~\ref{sec:applications} below.

\section{Generating QLCs Automatically} 
\label{sec:generating}

\subsection{A General Process for Creating QLCs}

\begin{figure}[tb]
  \centering
  \includegraphics[width=7cm]{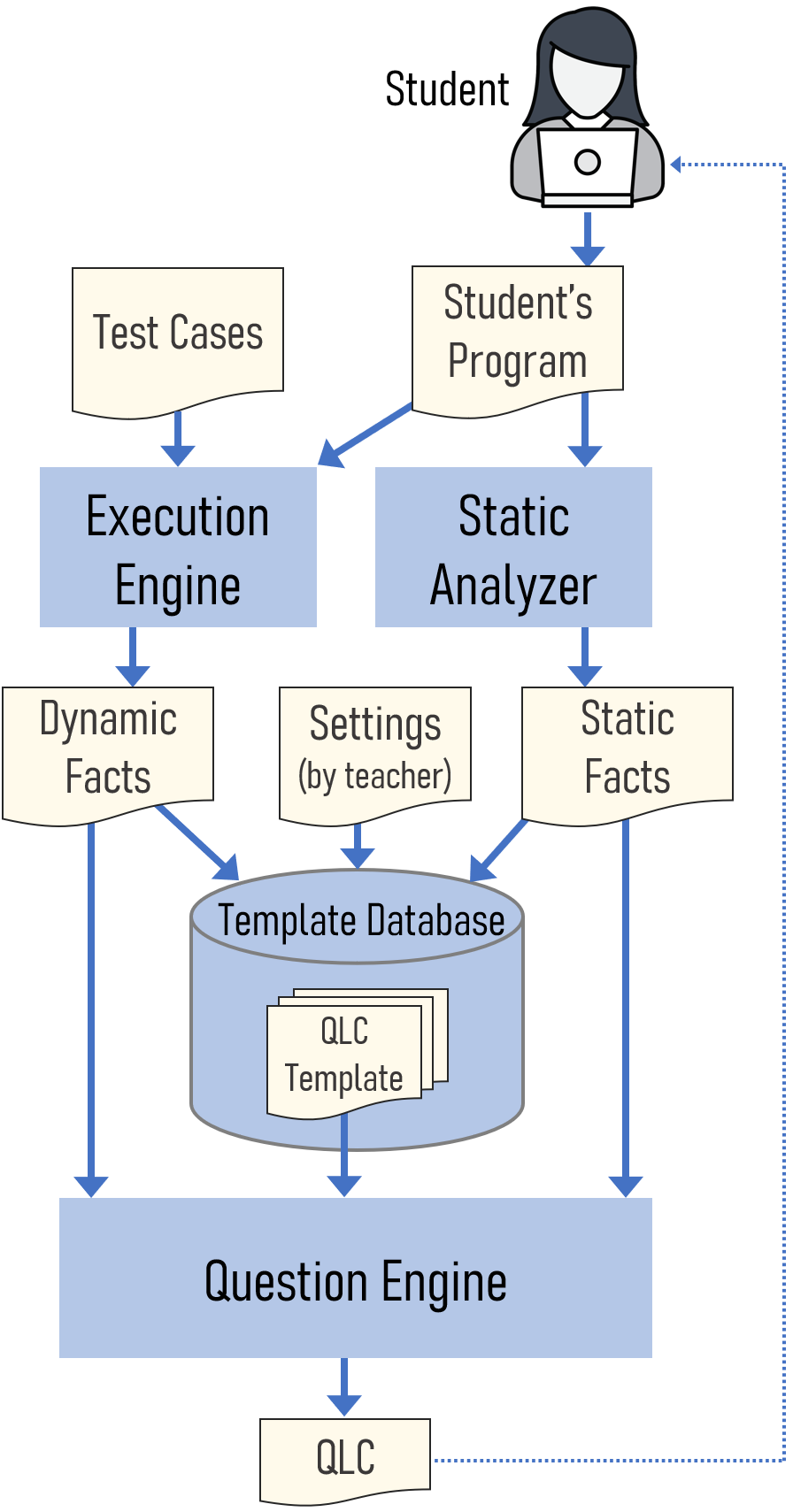}
  \caption{A process for generating QLCs.}
  \label{fig:arch}
\end{figure}

We propose a general process for generating QLCs, consisting of several steps (see Figure \ref{fig:arch}). 

The student’s program goes through a \emph{static analyzer} to extract \emph{static facts} from it, such as which variables there are and which functions are recursive. Static analysis of student programs has been successfully applied for purposes of generating feedback (e.g., \cite{Truong04}) and visualization (e.g., \cite{Santos2018EnhancingAnalysis}).
For QLCs, the static analysis does not need to be complex in order to be useful, as it may essentially consist of checking if certain structural elements are present in the student's program in order to decide if a QLC is applicable to it. More ambitious analyses may be possible, too, of course.

The program also goes through an \emph{execution engine} that collects \emph{dynamic facts} by executing the code; typically, this execution would be based on teacher-provided test cases. Dynamic facts involve variable values, control flow, and call stack history; for instance, the call stack depth at particular function call is a dynamic fact. (One implementation approach for the execution engine is to use a debugger interface if one is available for the programming language; cf.~\cite{Striewe2011UsingTutoring}.)

A \emph{template database} stores \emph{QLC templates}; see Table~\ref{tab:absquestions} for examples. In addition to the question texts shown in the table, each QLC template includes a list of requirements that a program must meet for the question to apply. For example, a question might only apply if the program features recursion. 
 
The static and dynamic facts---especially the former---are used to determine which \emph{QLC templates} are applicable. Many QLC templates are likely to match any given program; the selection of templates among the applicable ones can be further adjusted through teacher-governed settings. These settings may be specific to a course or even a particular programming assignment. They specify which kinds of QLCs are preferred and how many QLCs should be generated. Randomization can add further variety to the questions that are presented to different learners or to the same learner at different times.

The \emph{question engine} instantiates each chosen QLC template using the static and dynamic facts, thus producing a QLC (and its correct answer, if possible). The templates are instantiated to match the learner’s concrete program or part thereof (such as a function). For example, the QLC template \emph{``How deep does the call stack grow from executing function [F]?''} might be instantiated by replacing function [F] with {\small \texttt{smallest("ABBA")}}; similarly, the other templates in Table~\ref{tab:absquestions} would have the parts in square brackets filled using facts collected from the student's program.

\subsection{Other Inputs to QLC Generation}

By definition, learners’ code is the key input to QLC generation. However, there are other sources of information that might be additionally exploited in order to generate better QLCs. 

For example, the results of passed and failed unit tests might be used to tailor QLCs. Students could be asked about differences between their code and a model solution. Teachers might annotate model solutions, e.g., by attaching subgoal labels to sections of code; QLCs could then target those subgoals. (Some low-level, domain-generic subgoals such as ``initialize variables before starting loop'' might be detected automatically, without a teacher's intervention.) Students might be asked to compare their program to other programs which would also need to be provided as input to the QLC generator.

To personalize question topics and difficulty, a QLC generator would also need data on the learners’ prior coursework, including which QLCs the learners have previously answered.

\section{Use Cases for QLCs}
\label{sec:applications}

There are a number of scenarios where QLCs might usefully complement automatic assessment or otherwise integrate into programming education. We outline some of them below.

\subsection{Use Case: QLCs after Unit Tests}
\label{subsec:MainUseCase}

Automatic assessment systems collect students’ program code and execute it in a safe environment for analysis and feedback. These platforms could be extended with a QLC service that takes student submissions as input, generates QLCs, presents them to students, collects students’ answers, and provides feedback on those answers. 

In an automatic-assessment context, one option is to ask QLCs \emph{after successful testing}---i.e., once the student’s program passes functional tests. In this scenario, students are expected to have a good understanding of their own program when the QLCs are presented and the QLCs might serve as additional assessments. This is the use case that we are currently focusing our implementation efforts on. 

\begin{figure}[tb]
  \centering
  \includegraphics[width=7cm]{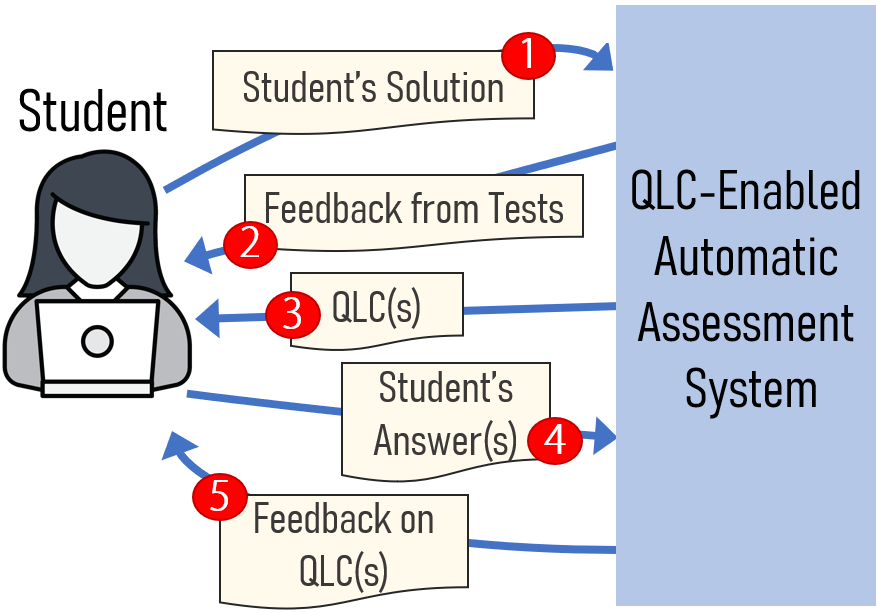}
  \caption{QLCs integrated with automated assessment. The numbers indicate the timeline of the interaction between a student and the system.}
  \label{fig:approach}
\end{figure}

Figure~\ref{fig:approach} illustrates the planned workflow between the learner and the system in this scenario. The learner writes a solution for a given exercise served by the system, which holds unit tests for it. After submitting a program for assessment (1), the learner receives the test results as feedback (2). Upon passing the tests, the learner is additionally presented with one or more questions about their solution (3). The learner answers the questions (4) and immediately receives feedback on the answers (5). That feedback, like the questions, is phrased in terms of the student’s own code.

\subsection{Alternative Use Cases}
\label{subsec:AlternativeUseCases}

Alternatively or additionally, an automatic assessment system could ask QLCs \emph{after failed tests}. This may prompt reflection but could also distract the student inconveniently unless the questions are particularly helpful. 

An automatic assessment system might also ask QLCs \emph{before testing}, perhaps as a precondition to displaying test results. That timing might be perceived as intrusive by students, however; moreover, the literature suggests that self-explanation prompts tend to be more useful when the student already knows whether the work they are explaining is correct or incorrect~\cite{RittleJohnson:ConstraintsOnSelfExpl}, and the same may apply to QLCs. 

A different setting for QLCs is the learners’ programming environment, such as an IDE. If the students run unit tests in this environment, QLCs might be presented before or after those tests, as described for automatic assessment systems above. Alternatively, students might be given the opportunity to ask for QLCs when they wish to answer them. A QLC-enabled IDE might even notify students when a potentially interesting QLC is available after a change to the program, although this option would need an exceptionally successful implementation so as to be both useful and unintrusive. 

Interactive ebooks with embedded programming assignments are increasingly common (e.g.,~\cite{Ericson:RunestoneAsEbookPlatform, Korhonen:EbookWG, Sirkia:PVsInEbook}). They offer similar integration options for QLCs as automatic assessment systems and IDEs do.

\subsection{Alternative Forms of QLCs}

In this paper, most of our discussion revolves around the notion of generating questions about a single program at a time. However, asking questions that link multiple programs also merits consideration. A student could be asked to comment on the commonalities or differences between their own program and another, which other program could have been authored by the teacher, the student themselves, or even another student. An example of such a cross-programmatic question appears at the bottom of Table~\ref{tab:absquestions}. To promote transferable learning of problem-solving schemata, these questions could ask students to identify recurring patterns such as roles of variables~\cite{Sajaniemi2005AnProgramming} or labeled subgoals~\cite{Margulieux:TeachingSelfExpl, Morrison:CuriousLoops}.

We have also focused our discussion on programming assignments that are closed-ended in the sense that they have a well-defined solutions that can be automatically assessed with predefined unit tests. QLCs could be generated from open-ended student code as well. Generating QLCs from open-ended code is simpler if the questions depend on static facts only, but QLCs with a dynamic nature may also be realistic if, for instance, the student has created unit tests.

\section{Challenges and Limitations}

Assessments and self-explanation prompts direct students’ attention to what is being asked and divert attention from other content that may be as important or even more so~\cite{RittleJohnson:ConstraintsOnSelfExpl}. Whether QLCs are used as assessment items or SEPs, this presumably applies to them as well. This represents a technical and pedagogical challenge: we must generate questions that are relevant to the appropriate learning objectives. Having the technical ability to ask about a property of a program does not mean that it is appropriate to pose that question to students; whether a question is appropriate varies and depends on context and goals. Moreover, even where a QLC \emph{is} appropriate for a particular situation, we must be alert to the possibility that its presence may de-emphasize other content that is also important but was not asked about.  

Whether a QLC is good also depends on the learner’s prior knowledge of programming and comprehension of their own program~\cite{corno1986adapting,kalyuga2003}. In cases where students \emph{do} understand the code that they have written, they may find QLCs tiresome. If QLCs are forced on students during problem solving, they may annoy the students. QLCs posed after passing unit tests may be perceived as a nuisance, as students may feel that getting the program to work is alone a sufficient assessment of their ability and that they have already done the hard part; ``selling'' the idea of QLCs to students appropriately in pedagogy may be key. 

To avoid unsuitable and repeated questions, it may be necessary to track learners’ history of programming assignments and QLCs. Teacher settings could define policies so that a particular QLC templates stops being used once a student has answered such questions correctly a set number of times.

Research suggests that simple multiple-choice questions are not ideal as self-explanation prompts~\cite{Bisra:SelfExplReview}. Open-ended QLCs can be more varied and richer in content than multiple-choice or single-value questions; on the other hand, open-ended questions are currently difficult or impossible to assess automatically. Progress is expected in automatic grading of short-text answers about programming~\cite{Fowler2021}.

Having reviewed the literature on self-explanation, Rittle-Johnson and Loehr~\cite{RittleJohnson:ConstraintsOnSelfExpl} note that ``the substantial time demands of self-explanation raises the question of when alternative activities would more easily or effectively achieve the same learning outcomes.'' Similarly, the additional time expenditure from answering QLCs must be evaluated against the questions’ possible benefits.

\section{Ongoing and Future Work}

As the QLC concept is very broad and flexible, there is a lot of ground to be covered in investigating the technical challenges of QLC generation, different types of QLCs, and different pedagogical uses of QLCs.

\subsection{Technical Development and Evaluations}

We are currently working on QLC generation in an automatic assessment system that poses questions to students after unit tests pass, as described in Section~\ref{subsec:MainUseCase}. This software will largely follow the question-generation process outlined above in Figure~\ref{fig:arch}, albeit initially it will have limited teacher configuration and template selection will be based on static facts only.

Using this implementation, we hope to evaluate: 

\begin{enumerate}
\item to what extent students are capable of answering QLCs; 
\item how students feel about QLCs;
\item if students' ability to answer QLCs correlates with other outcome measures; and
\item how all of the above depends on the type of QLCs.
\end{enumerate}

\noindent Early work along these lines is underway. Moreover, we are conducting a preliminary investigation where we pose \emph{manually created QLCs} to students after a programming assignment~\cite{lehtinen2021} (loosely similarly to our main scenario for automatic QLCs). In this pilot study, a system presents students with self-explanation questions that a researcher has tailored for the particular programming assignment and that are thus not chosen automatically or filled in based on facts collected from the student’s program.
The results from this pilot study have been tentatively promising so far: some students have reported that QLCs helped them to reflect on their program comprehension, while others noted that they had already thought about the QLC topics as they wrote their program. Furthermore, results confirm that many students indeed do have difficulties comprehending their own code and that the ability to explain one's own code correlates positively with later success in the course. These early findings highlight the potential of QLCs to improve program comprehension and its assessment.

\subsection{Further Opportunities}

Beyond our primary QLCs-after-passing-unit-tests scenario, there are many paths for future research, such as the following.

QLCs could be posed to students after failed unit tests, before testing, or even during the program-authoring process, as outlined in Section~\ref{subsec:AlternativeUseCases} above. 

Different ways of assessing QLCs could be explored: automatic, peer, self, and teacher assessment.
 
A possible thread of future work is to correlate students’ ability to answer QLCs with their self-explanation ability. This could provide insight into the mechanisms by which QLCs work (or fail to). Research could investigate whether students can be taught and motivated to use QLCs more effectively as self-explanation prompts and whether such teaching impacts on their QLCs responses and learning.

The impact of QLCs on plagiarism could be evaluated. Such research will need to weigh the possible benefits QLCs on plagiarism prevention against any negative consequences of mandatory and frequent QLCs.

Personalization of QLCs through learner modeling is an interesting and challenging area for future exploration. If the QLC generator could estimate each student's knowledge of different concepts, it could pose questions that suit the particular student better. For example, the questions might target possible misconceptions (analogously to~\cite{Gusukuma:MisconceptionDrivenFeedback}). Students' responses to QLCs could also be a source of data \emph{about} misconceptions (analogously to~\cite{Sirkia:ExploringVPSMistakes}). 

Finally, the techniques developed for generating QLCs might find applications in other contexts, such as generating practice questions from programs authored by teachers or peers.

\section{Conclusion}

We have proposed the idea of automatically generating questions, QLCs, about students’ program code for the students themselves to answer. This proposal opens up opportunities in deepening students’ program comprehension, complementing automatic assessment, and discouraging plagiarism. We have suggested several scenarios in which QLCs might be used and outlined the software support needed for generating QLCs; work on one such system is underway.

The practical feasibility and effectiveness of QLCs are as yet untested, but this proposal sets down a research agenda for future investigations. We have painted a fairly broad landscape in which computing education researchers may work on QLCs. In ongoing work, we are exploring a sector of that landscape; we hope that others might do similarly.

\bibliographystyle{IEEEtran}
\bibliography{references}

\end{document}